\documentclass[fleqn,10pt]{wlscirep}

\usepackage[utf8]{inputenc}
\usepackage[T1]{fontenc}
\usepackage{graphicx}
\usepackage{dcolumn}
\usepackage{bm}
\usepackage{amsmath,amssymb,bbm}
\usepackage[normalem]{ulem}
\usepackage{hyperref,float}
\usepackage{xcolor}
\usepackage{theorem}

\newcommand{\Np}{N_{\rm p}}
\newcommand{\Npt}{N_{{\rm p},t}}
\newcommand{\Nph}{\widehat{N}_{\rm p}}
\newcommand{\Npht}{\widehat N_{{\rm p},t}}
\newcommand{\kapht}{\widehat \kappa_{t}}
\newcommand{\kapt}{\kappa_{t}}

\newcommand{\vect}[1]{\boldsymbol{#1}}

{\theorembodyfont{\upshape}
\newtheorem{theo}{Definition}}

\title{The switching mechanisms of social network densification}

\author[1, 2]{Teruyoshi Kobayashi$^{\dagger}$}
\author[3, 4]{Mathieu G\'enois$^{\ddagger}$}
\affil[1]{Department of Economics, Kobe University, Kobe, Japan}
\affil[2]{Center for Computational Social Science, Kobe University, Kobe, Japan}
\affil[3]{CNRS, CPT, Aix Marseille Univ, Universit\'e de Toulon, Marseille, France}
\affil[4]{GESIS, Leibniz Institute for the Social Sciences, K\"oln/Mannheim, Germany}
\affil[$\dagger$]{kobayashi@econ.kobe-u.ac.jp}
\affil[$\ddagger$]{mathieu.genois@cpt.univ-mrs.fr}

\begin{abstract}
Densification and sparsification of temporal networks are attributed to two fundamental mechanisms: a change in the population in the system and/or a change in the chances that nodes in the system are connected. In theory, each of these mechanisms generates a distinctive type of densification scaling, but in reality both types are generally mixed.
Here, we develop a Bayesian statistical method to identify the extent to which each of these mechanisms is at play at a given point in time, taking the mixed densification scaling as input. 
We apply the method to networks of face-to-face interactions of individuals and reveal that the main mechanism that causes densification and sparsification occasionally switches, the frequency of which depending on the social context. The proposed method uncovers an inherent regime-switching property of network dynamics, which will provide a new insight into the mechanics behind evolving social interactions.

\end{abstract}

\begin{document}


\flushbottom
\maketitle
\section{Introduction}
Network representation of complex interactions among elements is an overarching framework  heavily used in many fields of science~\cite{Newman2010book,Barabasi2012NatPhys,Barabasi2016book}. 
For social systems, the dynamics of interactions between individuals (whether electronic, online or face-to-face) can be represented as time-varying networks, often called temporal networks, in which nodes come and go and edges are activated or deactivated as time goes on~\cite{Holme2012PhysRep,Holme2015}. Many essential features of human behaviour encoded in the representation of temporal networks have been revealed over the past decade, such as burstiness~\cite{Jo2011PlosOne,Karsai2011PhysRevE,Karsai2012SciRep}, circadian/diurnal rhythms~\cite{Jo2012NewJPhys}, temporal communities~\cite{gauvin2014detecting}, higher-order interactions~\cite{scholtes2016higher,lambiotte2019networks}, etc.  

While the studies of temporal networks shed light on the time-varying nature of interactions between nodes, dynamics in social systems emerge not only at the local level~\cite{Gautreau2009PNAS}, but also at the global level. 
In a wide variety of social contexts, network size (\emph{i.e.}, the number of active nodes) and the number of edges observed at a given point in time are very often not constant, and accordingly the average degree increases or decreases~\cite{Leskovec2007CA_Full,Kobayashi2020}. In fact, the numbers of aggregate nodes and edges have been shown to have a scaling relationship known as the densification power law or densification scaling~\cite{Leskovec2007CA_Full}.
 In temporal networks (\emph{i.e.}, a sequence of snapshot networks), any variation in the number of active nodes $N$ and the number of edges $M$ can be \emph{a priori} attributed to changes in (i) the population in the system (\emph{e.g.}, the number of students present in a school, the number of attendees in a conference, etc); (ii) the probability of two nodes being connected; or (iii) both. With a constant probability of edge creation, $N$ and $M$ will increase if more nodes enter the system, since each node will have a higher chance of finding partners. Likewise, for a given population, if the probability of two nodes being connected increases, $M$ will surely increase, and $N$ will rise as well as isolated nodes, if they exist, will be more likely to get connected.

These two mechanisms are fundamental factors that bring about the dynamics of $N$ and $M$, yet separating their contributions based on the dynamical behaviour of $N$ and $M$ is a challenging problem. In a wide variety of social and economic systems, network dynamics are likely to be driven by a mixture of these two mechanisms, and moreover their relative importance may occasionally change as the network evolves~\cite{Kobayashi2020}. 
In theory, each of these two mechanisms leads to a distinctive type of densification scaling; The first one, generated by the evolution of population, is a scaling behaviour similar to the typical densification scaling in which the number of edges $M$ scales with the number of active nodes $N$ with a constant exponent $\alpha$, \emph{i.e.}, $M\propto N^\alpha$~\cite{Leskovec2007CA_Full}. The second one is an accelerating growth of $M$, which is caused by the evolution of the probability of edge creation~\cite{Kobayashi2020}.
In fact, for the human contact networks we study, neither of these two types of scaling is observed in their original form. Rather, we observe a ``mixed'' scaling behaviour which appears to be a composite of the two types and therefore cannot be explained by a single scaling law. 

Here, we develop a Bayesian statistical method to identify the source of dynamics generating network densification and sparsification based on the sequence of $N$ and $M$. To take into account possible changes in the source of dynamics, we derive two specifications (\emph{i.e.}, ``regimes'') for the solution of a simple generative model, namely a dynamic hidden variable model, each of which capturing one of the two fundamental mechanisms. By fitting the two specifications simultaneously to the observed mixed scaling relationship using a unified estimation framework, known as the Markov regime-switching model~\cite{Hamilton1994book,hamilton2010regime}, we are able to estimate the probability that the dynamical source of densification or sparsification at a given point in time is attributed to a particular mechanism. At the same time, the Bayesian inference also allows us to trace the paths of the time-varying parameters directly related to the dynamical source, \emph{i.e.}, the population in the system and the activity level of nodes.
An important advantage of the regime-switching model is that it allows the ``true'' model specification to occasionally switch, possibly depending on the social context.

In this work we analyse networks of face-to-face human interactions collected by the SocioPatterns collaboration\cite{SocioPatterns}. We focus on four datasets: contact networks in two scientific conferences, a hospital and a workplace. Such networks can indeed be affected by the two fundamental mechanisms at the same time, because (i) individuals can always enter and exit the system, and (ii) presence of a time schedule could facilitate or inhibit face-to-face interactions (\emph{e.g.}, attendees of a conference are more likely to have interactions during coffee breaks than during keynote talks). In particular, using data on academic conferences has an important advantage, as it allows us to compare the dynamical regimes detected by the proposed method with the ``ground-truth'' conference time schedules.
We find indeed that during keynote talks, parallel sessions and coffee breaks, the temporal densification and sparsification in the contact networks formed by conference attendees are mainly related to shifts in the chance of contacts being made between attendees present at the venue. On the other hand, shifts in the population are the main driving force of densification and sparsification during registration and poster sessions. This result is consistent with our intuition that the number of attendees in the middle of the program would be mostly constant, while it may be more likely to change during registration, which is held in the morning, and poster sessions in which not all of the attendees participate. For contact networks in a hospital and a workplace, this kind of comparison with a prespecified time schedule is not possible because there is no such rigorous time constraints to follow. Nevertheless, in all the systems we examined, the proposed method reveals that the main driving force of network densification and sparsification is occasionally switching, suggesting that the formation of social ties in physical space generally involves multiple dynamical sources.

\section{Results}

\subsection{Empirical evidence on mixed densification scaling}

\begin{figure*}[tb]
    \centering
    \includegraphics[width=17.5cm]{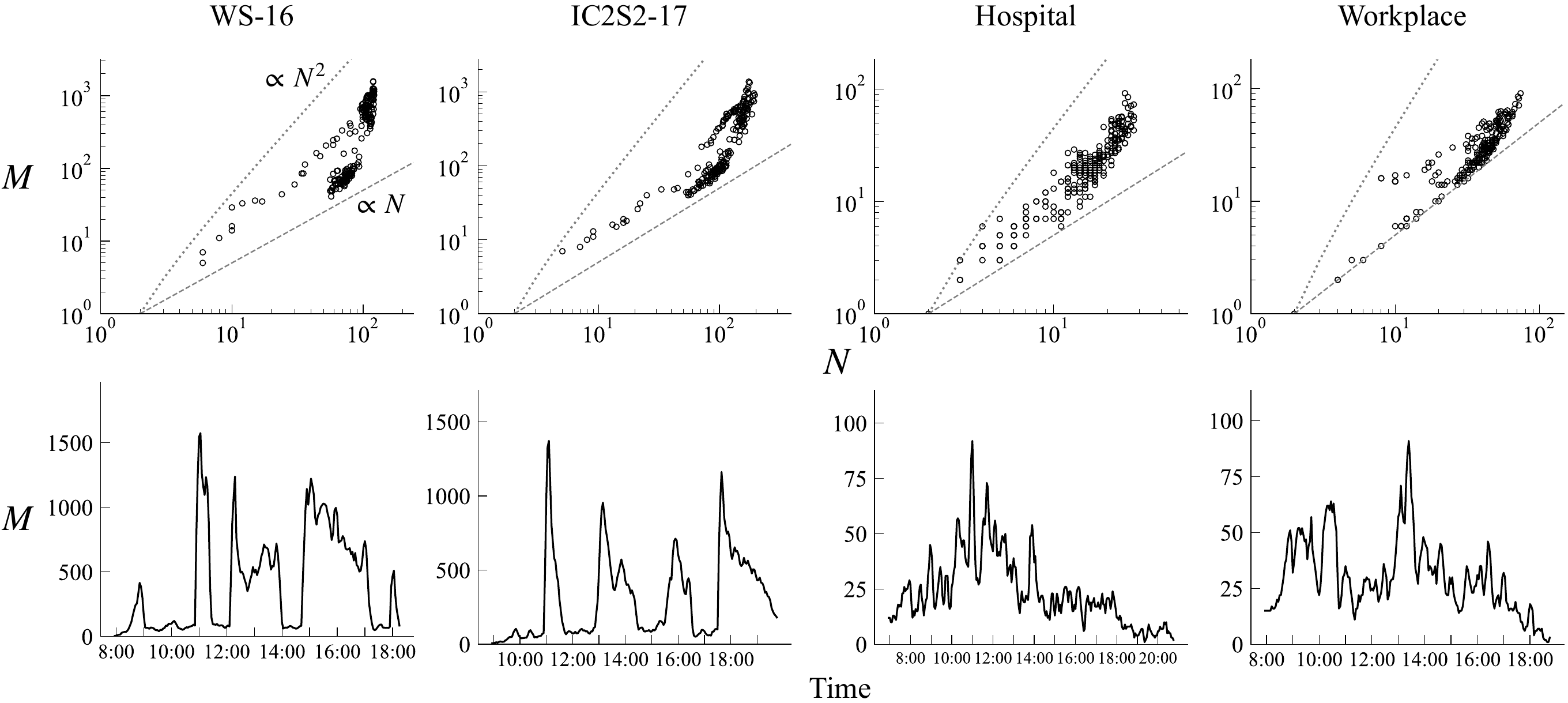}
    \caption{Dynamical behaviour of number of active nodes $N$ and number of active edges $M$. In upper panels, dynamical relationship between $N$ and $M$ is shown. Each dot represents a snapshot network created over a 10-minute time window. Gray dashed and dotted lines respectively denote $N/2$ (\emph{i.e.}, the lower bound for $M$) and $N(N-1)/2$ (\emph{i.e.}, the upper bound for $M$). Lower panels show the behaviour of $M$ over time.}
    \label{fig:scaling_data}
\end{figure*}

We focus our analysis on temporal contact networks taken from the following four datasets:
\begin{itemize}
    \item {\bf{WS-16}}: Contacts between participants of the Computational Social Science Winter Symposium 2016 at GESIS in Cologne on November 30, 2016~\cite{Genois2019}. 
    \item {\bf{IC2S2-17}}: Contacts between participants of the International Conference on Computational Social Science 2017 at GESIS in Cologne on July 12, 2017~\cite{Genois2019}.
    \item {\bf{Hospital}}: Contacts among patients, nurses, doctors and staffs in a Hospital in Lyon on December 8, 2010~\cite{Vanhems:2013}.
    \item {\bf{Workplace}}: Contacts between workers in a office building in France on June 27, 2015~\cite{genois2015data}.
\end{itemize}
These data consist of contacts between individuals collected every 20 seconds using RFID sensors \cite{Cattuto2010PLOS,SocioPatterns}. A ``contact'' is here defined as a physical, face-to-face proximity event. The datasets thus give us temporal networks in which nodes are individuals and edges encode the contacts occurring between them.
All datasets exhibit large and abrupt fluctuations of the number of edges that are typical in these non-stationary systems (see Fig.~\ref{fig:scaling_data}, lower panels). In these particular contexts of social interactions, these transitions between high an low activity periods are often related to specified schedules: from talk sessions to coffee breaks in the conferences, changes in shifts in the hospital, from desk work to meetings in the workplace.

In many social and economic dynamical networks, the numbers of aggregate edges and nodes have a superlinear scaling relationship called the ``densification power law''~\cite{Leskovec2005HepAS,Leskovec2007CA_Full,bettencourt2009scientific}, in which the average degree is increasing with the number of nodes, \emph{i.e.}, ``densification''. For temporal networks, where there is a sequence of network snapshots, a similar type of scaling emerges from the dynamics of the population, in which nodes enter and leave the system, keeping the chance of two nodes being connected constant~\cite{kobayashi2018social,Kobayashi2020}. However, another type of scaling emerges in real-world systems for which the population is fixed. In such systems densification is ``explosive'', with the scaling exponent increasing with $N$\cite{Kobayashi2020}.
While these two classes of scaling could be differentiated and identified from data if we observe a specific type of scaling~\cite{Kobayashi2020}, in general there may exist a mixture of them that cannot be easily classified as one of the two classes.
Indeed, in the four datasets we study, no clear scaling relationship appears (Fig.~\ref{fig:scaling_data}, upper panels).
In the following we show that the mixed shape of empirical densification behavior reflects a mixing of both classes of scaling.

\subsection{Two dynamical regimes in the dynamic hidden-variable model}
To explore the temporal dynamics of densification and sparsification, we consider a dynamic version of the hidden variable model. The probability that two nodes $i$ and $j$ are in contact within a given time window $t$ is:
\begin{align}
    \mathcal{P}_{ij,t} = \kappa_t  a_{i} a_{j}, \;\;\; i,j=1,\ldots, \Npt, \;\; t = 1,\ldots, T.
    \label{eq:prob_ij}
\end{align}
where $a_{i}$ is the ``fitness'' that represents the intrinsic activity level of node $i$~\cite{Caldarelli2002PRL,Boguna2003PRE,DeMasi2006PRE}, and $T$ denotes the last time window in the data. There are two time-varying parameters in the model. The first one is $\kappa_{t}>0$, which modulates the overall activity rhythm of nodes. A variation in $\kappa$ would reflect the time-schedule of a conference or a school, working hours in an office or a hospital, or the circadian rhythm of individuals~\cite{Cattuto2010PLOS,Jo2012NewJPhys,aledavood2015digital,kobayashi2019structured}. The second time-varying parameter $\Npt$ denotes the potential number of active nodes at time $t$, \emph{i.e.}, the total of active and inactive nodes that are in the room or the building.
It should be noted that although the number of active nodes (\emph{i.e.}, nodes having at least one edge) $N_t$ is always observable from the data, the potential number of nodes $\Npt$ is not. We do not usually know how many people were actually in the room at a given time because people could enter and exit the room at any time without being interacting with any other individual. We can observe the number of active nodes that appear in the record of contacts, but in many cases there is no record of nodes without any interaction.
We assume that activity $a_i$ is uniformly distributed on $[0,1]$, because i) we do not have any prior information about the full distribution of the activity levels of all nodes including isolated ones, and ii) introducing a more general distribution prohibits us from obtaining an analytical solution, which makes it difficult to implement parameter estimation.

The average numbers of active nodes $N$ and edges $M$ are analytically given as (see section~\ref{sec:analytical_derivation_N_M} in Methods for derivation):
\begin{align}
N &= \Np \left[ 1-  \frac{2}{\kappa\Np}\left(1-\left( 1-\frac{\kappa}{2}\right)^{\Np}\right) \right],\label{eq:N_main}\\
M &= \frac{1}{8} \kappa\Np(\Np-1),\label{eq:M_main}
\end{align}
where we drop time subscript $t$ for brevity. From these expressions, it is clear that the two parameters $\kappa$ and $\Np$ play different roles in the determination of $N$ and $M$, but it is not clear how $N$ and $M$ correlate. To see the direct relationship between $N$ and $M$, we eliminate one of the two parameters in Eq.~\eqref{eq:N_main}, using Eq.~\eqref{eq:M_main}. By doing this, we can effectively endogenise either $\kappa$ or $\Np$.
Depending on whether we endogenise $\kappa$ or $\Np$, we obtain different functional forms that connect $N$ and $M$.

\subsubsection{Regime 1: $\Np$-driven dynamics}
First, let us consider the case of time-varying $\Np$. This is a situation in which the dynamics of $N$ and $M$ are fully driven by changes in the population. We call this system as being in ``Regime 1'' or ``state 1'':
\begin{theo} {A system is in Regime 1 if $\Np$ is time-varying and $\kappa$ is constant, in which case the dynamical relationship between $N$ and $M$ is given by:}
\begin{align}
    N_t &= \Np(M_t,\kappa) \left[ 1-  \frac{2}{\kappa\Np(M_t,\kappa)}\left(1-\left( 1-\frac{\kappa}{2}\right)^{\Np(M_t,\kappa)}\right) \right] \notag \\
      & \equiv h^1(M_t;\kappa),
      \label{eq:model1}
\end{align}
where the time-varying $\Np$ value is expressed as a function of $M_t$ and $\kappa$: $\Np(M_t,\kappa) \equiv \frac{1+ \sqrt{1+{32M_t/{\kappa}}}}{2}$ (see, Eq.~\ref{eq:M_main}). 
\end{theo}

\begin{figure*}[tb]
    \centering
    \includegraphics[width=17.5cm]{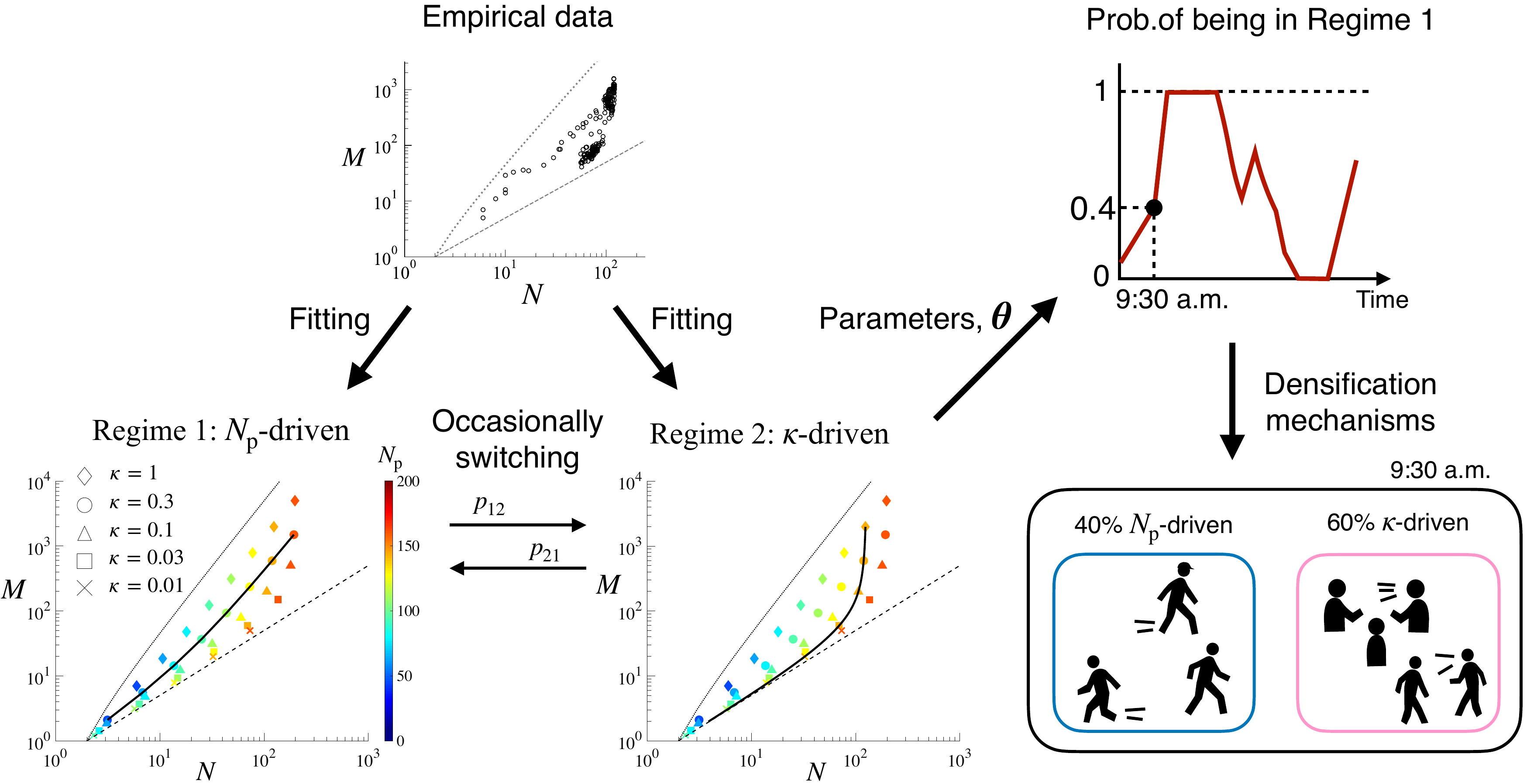}
    \caption{Schematic of the identification method. Empirical densification is fitted to the regime switching model in which the model switches from Regime 1 to Regime 2 (resp. from Regime 2 to Regime 1) with probability $p_{12}$ (resp. $p_{21}$). Then, the estimated parameters are used to infer the probability of the system being in Regime 1 at a given time $t$. For panels at lower left and lower middle, different colours denote different $\Np$, and different symbols denote different $\kappa$ (see Eqs.~\ref{eq:N_main} and \ref{eq:M_main}). If the scaling is $\Np$-driven (resp. $\kappa$-driven), the time variation of $N$ and $M$ is fully caused by shifts in $\Np$ (resp. $\kappa$).}
    \label{fig:schematic}
\end{figure*}

For the purpose of parameter estimation, we introduce an error term as
    $N_t=h^1(M_t;\widehat\kappa) + \varepsilon_{1,t}, \label{eq:h1}$
where $\widehat{\kappa}$ denotes the estimated value of $\kappa$, and $\varepsilon_t^1$ is a residual term following a normal distribution with mean zero and standard deviation $\sigma_1$.
Estimated value of $\Npt$ when the system is in Regime 1 leads to:
\begin{equation}
    \widehat N_{{\rm p},t}|_{S_t=1} = \frac{1+ \sqrt{1+{32M_{t}/{\widehat\kappa}}}}{2},
\end{equation}
where $S_t=1$ denotes the fact that the system is in Regime 1 at time $t$.
In Regime 1, network dynamics is totally driven by the time-varying nature of the population, what we call ``$\Np$-driven'' dynamics. For a given $\kappa$, the slope of densification scaling is close to constant, while different $\kappa$ yield different slopes. (Fig.~\ref{fig:schematic}, lower left).

\subsubsection{Regime 2: $\kappa$-driven dynamics}
Next, let us consider the case of time-varying $\kappa$. This corresponds to a situation in which the dynamics of the system is fully driven by changes in the overall activity of individuals. We call this system as being in ``Regime 2'' or ``state 2'': 

\begin{theo} {A system is in Regime 2 if $\kappa$ is time-varying and $\Np$ is constant, in which case the dynamical relationship between $N$ and $M$ is given by}
\begin{align}
    N &= \Np \left[ 1-  \frac{2}{\kappa(M,\Np)\Np}\left(1-\left( 1-\frac{\kappa(M,\Np)}{2}\right)^{\Np}\right) \right] \notag \\
      & \equiv  h^2(M;\Np),
      \label{eq:model2}
\end{align}
where the time-varying value of $\kappa$ is expressed as a function of $M_t$ and $\Np$: $\kappa(M_t,\Np) \equiv \frac{8M_t}{\Np(\Np-1)}$ (see, Eq.~\ref{eq:M_main}).
\end{theo}

For estimating, we add an error term as $N_t=h^2(M_t;\Nph) + \varepsilon_{2,t}, \label{eq:h2}$
where $\Nph$ denotes the estimated value of $\Np$, and $\varepsilon_{2,t}$ is a residual term following a normal distribution with mean zero and standard deviation $\sigma_2$. 
Estimated value of $\kappa$ at time $t$ when the system is in Regime 2 leads to:
\begin{equation}
    \widehat\kappa_t|_{S_t=2} = \frac{8M_t}{\Nph(\Nph-1)}.
\end{equation}
In Regime 2, network dynamics is fully driven by the individuals' time-varying activity levels, what we call ``$\kappa$-driven'' dynamics, and the slope of densification scaling in fact increases with $N$ (Fig.~\ref{fig:schematic}, lower middle). This kind of accelerating growth of $M$ naturally happens when edges are created in a fixed-population system, in which case the network tends to be denser as the number of inactive nodes vanishes.

\subsection{Analysis of switching dynamics behind temporal densification and sparsification}

\subsubsection{A Markov regime switching model}

In real-world networks, the mechanism of densification and sparsification may occasionally change depending on the context, such as working schedule, coffee breaks, lunch time, etc. To incorporate such a possibility, we propose a unified framework based on the Markov regime switching model in which the hidden state of a system can switch from Regime 1 to Regime 2 (respectively from Regime 2 to Regime 1) with probability $p_{12}$ (resp. $p_{21}$)~\cite{Hamilton1994book,hamilton2010regime}.
An important advantage of the regime switching model is that it allows us to calculate the probability of a system being in Regime $s\in\{1,2\}$ at time $t$ for a given parameter set $\vect{\theta} =\{\Np,\kappa,\sigma_1,\sigma_2,p_{11},p_{22}\}$.  
This probability of the system being in Regime $s$ can then be interpreted as the relevancy of each mechanism in explaining the densification dynamics at a given time (Fig.~\ref{fig:schematic}). 
We employ a Bayesian approach for the estimation of the parameters, using the Markov chain Monte Carlo (MCMC) to obtain posterior distributions (see, Methods~\ref{sec:method_regime_switch} for the estimation method).

In the following, we use the smoothed probability ${\rm Pr}(S_t=s|\psi_T;\vect{\theta})$ which is calculated conditional on all the information available at time $T$, denoted by $\psi_T$ (see, Methods~\ref{sec:smoothed} for full derivation)~\cite{kim1994smoothed}. Validation analyses using synthetic networks show that the proposed method correctly detects the switching of regimes and estimates the model parameters quite accurately (Table~\ref{tab:validation_params}, Figs.~\ref{fig:validation_prob_scatter} and \ref{fig:validation_Np_kap} in Supporting Information (SI)).
Given the probability of being in Regime $s\in\{1,2\}$, we can estimate the dynamical parameters $\Npt$ and $\kapt$ as:
\begin{align}
    \Npht &= {\rm Pr}(S_t=1|\psi_{T};\widehat{\vect{\theta}})\cdot\widehat N_{{\rm p},t}|_{S_t=1} \; +\;  {\rm Pr}(S_t=2|\psi_{T};\widehat{\vect{\theta}})\cdot\Nph, \\
    \kapht &= {\rm Pr}(S_t=1|\psi_{T};\widehat{\vect{\theta}})\cdot\widehat{\kappa} \; + \; {\rm Pr}(S_t=2|\psi_{T};\widehat{\vect{\theta}})\cdot\widehat\kappa_t|_{S_t=2},
\end{align}
where $\widehat{\vect{\theta}}$ denotes the set of estimated parameters, which is summarised in Table.~\ref{tab:parameters} in Methods.

\subsubsection{Classification of network dynamics}

\begin{figure*}[tb]
    \centering
    \includegraphics[width=17.5cm]{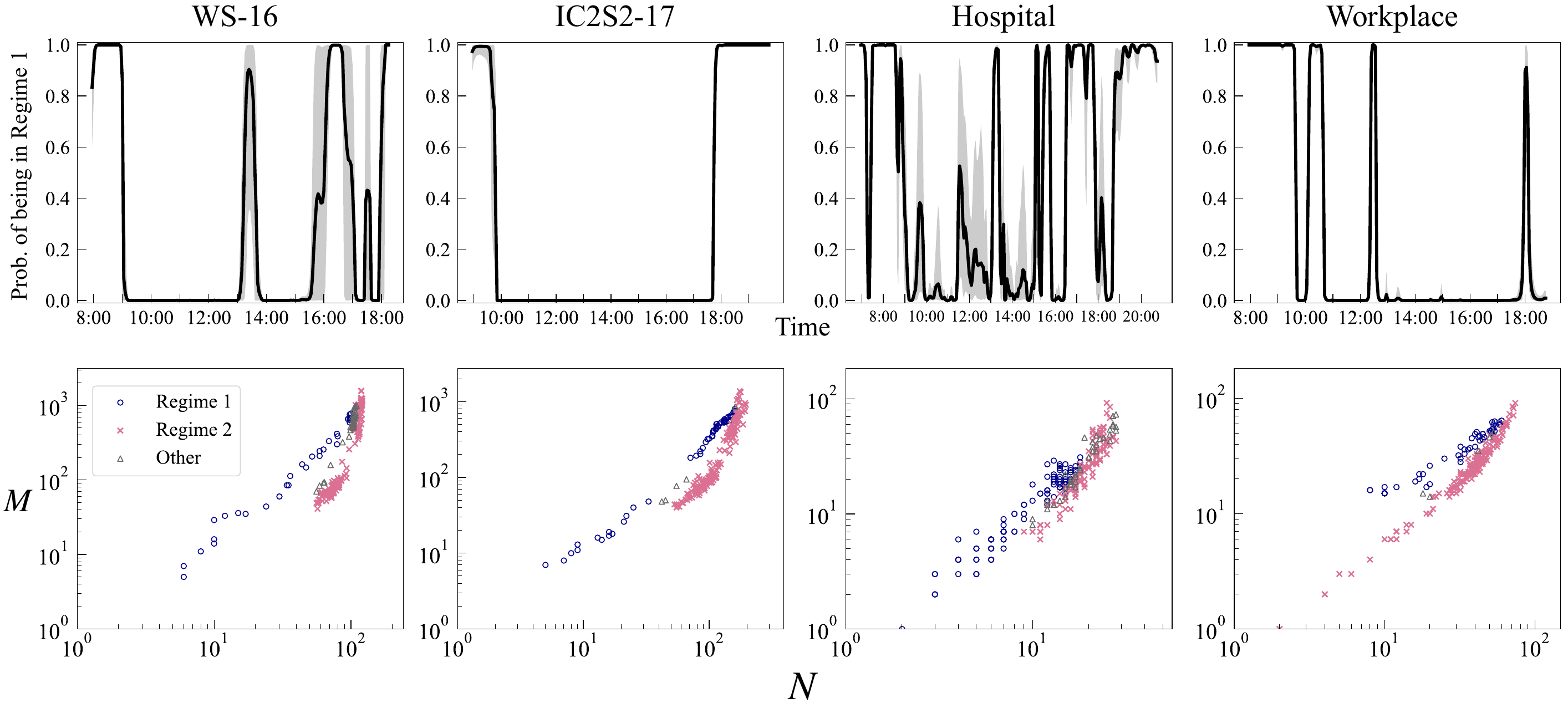}
    \caption{Identification of dynamical regime. Upper panels show the smoothed probability of being in Regime 1 (\emph{i.e.}, $\Np$-driven dynamics) at each time window. 95\,\% credible interval is indicated by shading. Lower panels show $N$-$M$ plots with classified regimes being denoted by different colours and symbols. We identify a snapshot network as being in Regime 1 (resp. Regime 2) if the estimated probability of being in Regime 1 (resp. Regime 2) is greater than 0.5 in more than 95\,\% of MCMC sampling. Otherwise, a network is considered as being in an undetermined ``gray area''.}
    \label{fig:regime_NM_result}
\end{figure*}

The Bayesian estimation of the parameters suggests that the empirical systems' dynamics are indeed occasionally switching between $\Np$-driven and $\kappa$-driven (Fig.~\ref{fig:regime_NM_result}, upper panels).
For the conference data, a common feature is that the probability of being in Regime 1 is almost 1 prior to the first session and after the last keynote session of the day, and mostly zero in between (see Fig.~\ref{fig:schedule} for the correspondence between the dynamics and the schedule of the conferences). For WS-16, we see further fluctuations between the two regimes, one linked to the lunch break, the other to the poster session which closed the day. This suggests that the dynamics during the oral sessions, keynote talks and breaks are mainly driven by changes in the activity level of participants, while in the ``opened'' time slots, such as registration, closing and poster session, their dynamics are explained by time-varying population. The same patterns linked to the schedule are found on the other days of the conferences (see \ref{fig:SI_regime_NM_result}a--c).

For the Workplace data we see a roughly similar pattern (Figs.~\ref{fig:regime_NM_result}, top right). The dynamics in the early morning and evening are driven by a variation in $\Np$, as well as around lunch time and coffee break, and changes in activity level are the main source of dynamics in between. This is, of course, not necessarily a general property of contact networks in physical space. We also see that the regime remains almost constant in most of the day (Fig.~\ref{fig:SI_regime_NM_result}e), or there might be days in which the regime constantly changes throughout the day (Fig.~\ref{fig:SI_regime_NM_result}f).
In the case of the Hospital data, there is no clear tendency for the regime-switching pattern (Figs.~\ref{fig:regime_NM_result}, third column and \ref{fig:SI_regime_NM_result}d), which seems natural for such an open environment with visitors and medical workers coming and going, and no general, fixed schedule for working hours.

We next attempt to classify the snapshot networks into two groups based on their probability of being in a particular regime. We identify a snapshot network at $t$ as being in Regime 1 (resp. Regime 2) if more than 95\,\% of samples for the value of ${\rm Pr}(S_t=1|\psi_T;\vect{\theta})$ generated by MCMC are greater than 0.5 (resp. lower than 0.5), \emph{i.e.,} in more than 95\,\% of parameter sampling the dynamics at $t$ is considered to be attributed to Regime 1 (resp. Regime 2). Otherwise, the system is considered as being in an undetermined ``gray area''.
As seen in the lower panels of Fig.~\ref{fig:regime_NM_result}, the location of snapshot networks in the $N$-$M$ space is strongly related to which regimes they belong to. As expected, the snapshots in Regime 1 exhibit a scaling whose slope is almost constant (\emph{i.e.}, $\Np$-driven scaling), while the snapshots in Regime 2 exhibit accelerating growth patterns (\emph{i.e.}, $\kappa$-driven scaling). Classifying each time window according to the underlying dynamical mechanism is essentially equivalent to identifying patterns in the $N$-$M$ space.

\subsubsection{Temporal dynamics of population and activity level}

\begin{figure*}[tb]
    \centering
    \includegraphics[width=17.5cm]{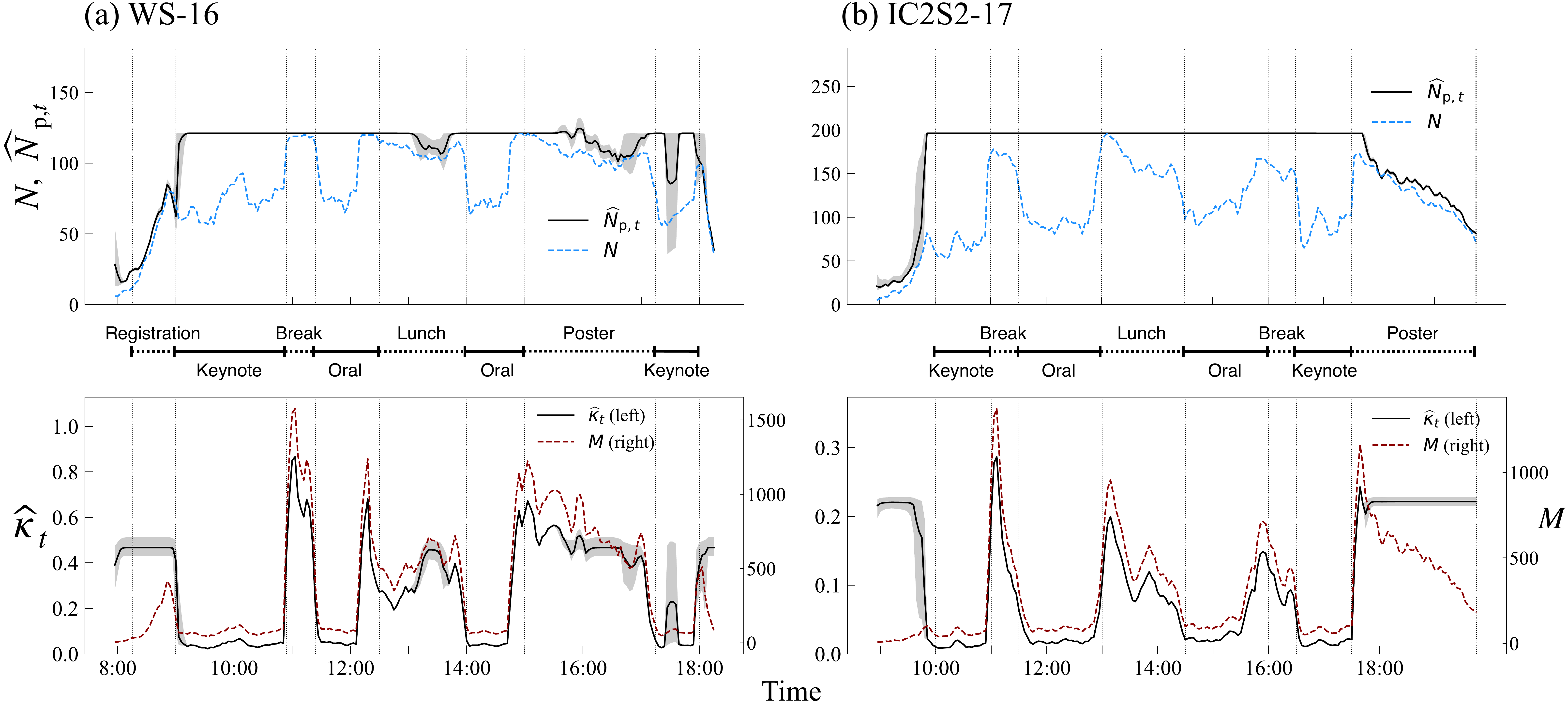}
    \caption{Estimation of $\Npt$ and $\kappa_t$ for (a) WS-16 and (b) IC2S2-17. $\Npht$ and $\kapht$ are shown in the upper and the lower panels, respectively, and the 95\,\% credible interval is indicated by shading. Upper panels show the number of active nodes (dashed blue line) at each time, thus the difference between the two lines represents the number of isolated nodes. Lower panels also show the number of edges at each time (dashed red line). Vertical dotted lines indicate the time windows of the scheduled sessions, with the labels in the middle.}
    \label{fig:schedule}
\end{figure*}

We also examine the evolution of the dynamical parameters for both regimes (Fig.~\ref{fig:schedule}).
For the two conferences (WS-16 and IC2S2-17), the estimated population size $\Npht$ increases at the beginning of the day and decreases at the end, consistent with the dynamics of participants entering and exiting the venue. The estimated activity parameter $\kapht$ is high during these periods, and the level is consistent with those seen in highly active windows during social breaks. During the main program, the population is virtually constant and the size is consistent with the number of attendants ($\sim$ 120 for WS-16, $\sim$ 200 for IC2S2-17). The variation of network size is thus mainly driven by the schedule, which constraints the participants' networking activity.
In the case of WS-16, the fluctuation of $\Npht$ during the lunch break and the poster session are worth noting since the variation of observed network size $N$ seems to be driven by both mechanisms; we see slight reductions in the estimated population while the overall activity is still high in these time windows. This demonstrates the ability of the proposed method to extract mixed-regime periods in which both of the two mechanisms are at work (see Fig.~\ref{fig:schematic}, right, for schematic). Similar patterns are also found in the other days (see~\ref{fig:SI_conf_schedule}).

\begin{figure*}[tb]
    \centering
    \includegraphics[width=17.5cm]{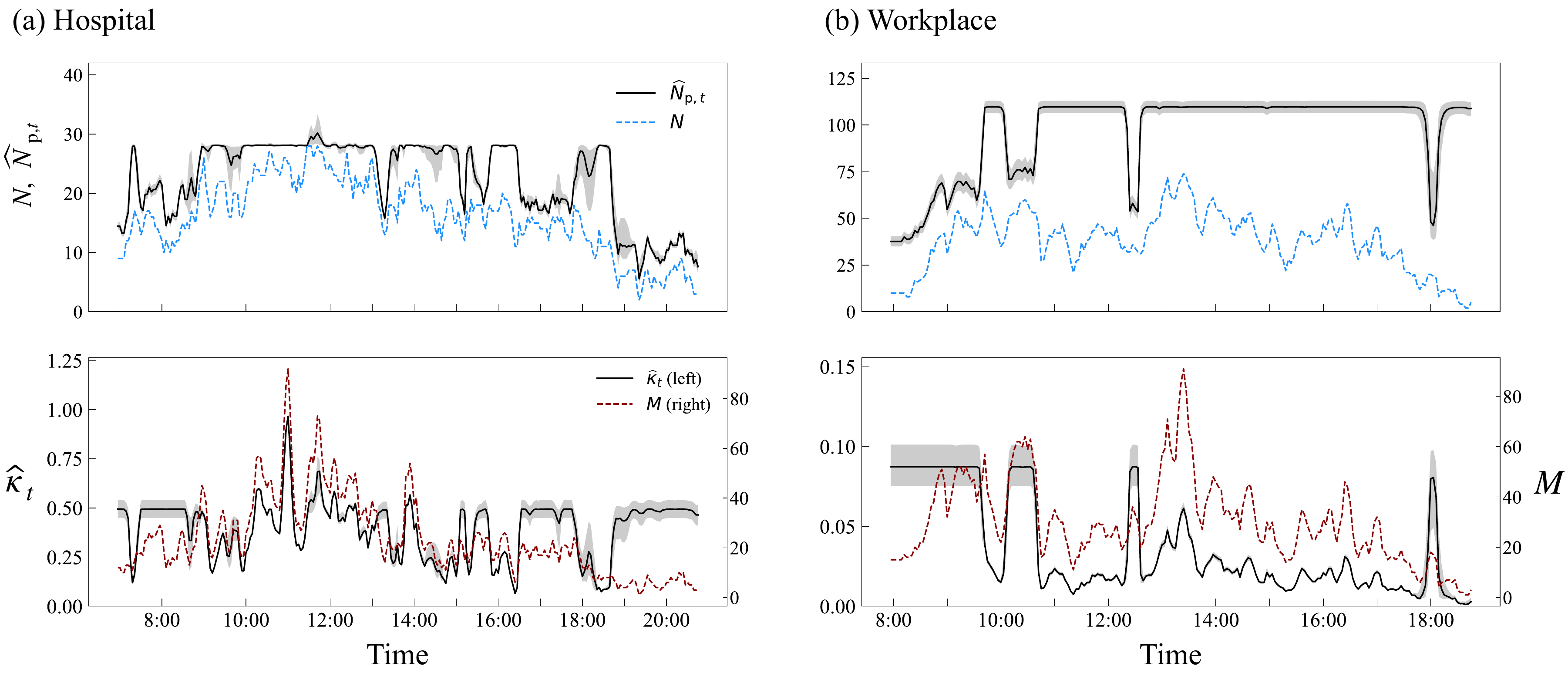}
    \caption{Estimation of $\Npt$ and $\kappa_t$ for (a) Hospital and (b) Workplace.}
    \label{fig:schedule_hos_work}
\end{figure*}

In the Hospital data, the regime-switching dynamics is much less periodic, with lots of transitions and mixed periods (Fig.~\ref{fig:schedule_hos_work}a).  This is however not surprising, because there is no fixed schedule regulating either the activity or the number of people present.
For the Workplace data, we also do not expect \emph{a priori} to see a clear segmentation of regimes because of the absence of a rigid schedule as in a hospital. However, the dynamics uncovered by our method indicates that the situation is much simpler than that for Hospital, as there seem to be less variation in population size, aside from the ``opening'' and ``closing'' effects and a reduction in population around the lunch time (Fig.~\ref{fig:schedule_hos_work}b). The day that exhibited many regime switches presents however many episodes of small variations in population size (see \ref{fig:SI_HosWork_schedule}), similar to the dynamics observed in a Hospital.

\subsubsection{Non-monotonic behaviour of network density}

\begin{figure*}[tb]
    \centering
    \includegraphics[width=17.5cm]{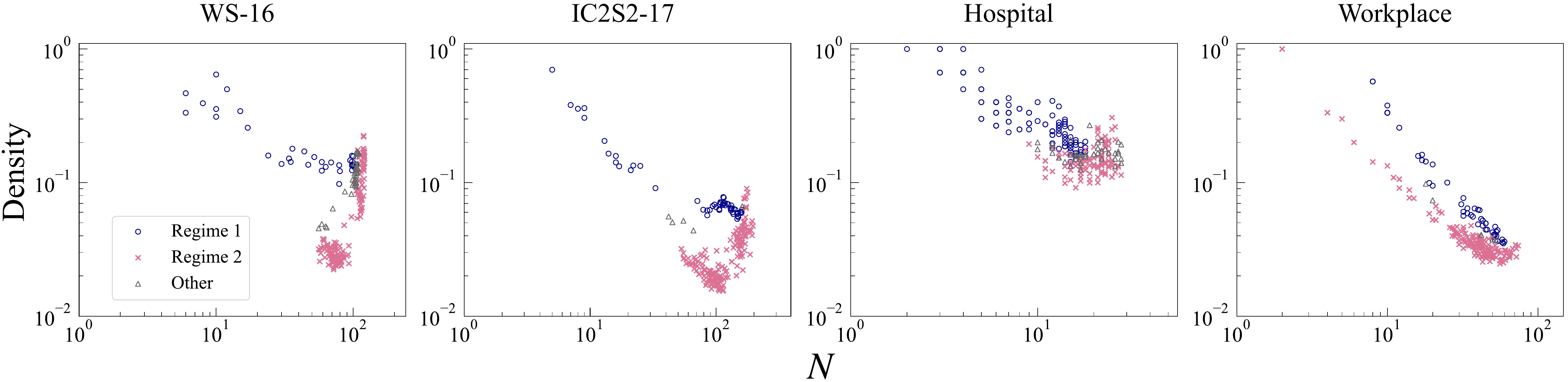}
    \caption{Density versus the number of active nodes. Classification of dynamical regimes is conducted in the same way as in Fig.~\ref{fig:regime_NM_result}.}
    \label{fig:density_main}
\end{figure*}

Since both types of scaling emerging from two different dynamics exhibit superlinearity, the average degree is always increasing in $N$. However, the density of networks, defined by $2M/(N(N-1))$, is not always increasing with $N$ (Fig.~\ref{fig:density_main}). In fact, when the dynamics is $\Np$-driven, the network density mostly decreases as the network size $N$ increases (Fig.~\ref{fig:density_main}, blue circle). So, a rise in $N$ causes the density to be smaller when the engine of dynamics is changes in population. In contrast, when changes in $\kappa$ play a dominant role, the network density may increase when the network size is sufficiently large (Fig.~\ref{fig:density_main}, pale-red cross). This is because when the number of active nodes $N$ is close to its upper bound $\Np$, at which the activity levels of remaining inactive nodes are fairly low, the overall activity $\kappa$ needs to be large enough for those low-activity nodes to get at least one edge. This would necessarily increase the total number of edges in the network to a large extent, which leads to a ``true'' densification of networks.

These properties are also confirmed by the analytical equation for the average network density~\cite{Kobayashi2020}  
\begin{align}
    \frac{2M}{N(N-1)} = \frac{\kappa}{4} \left( \frac{1}{1-q_0(\kappa,{N}_{\rm p})}\right)^2 \left( 1+\frac{q_0(\kappa,{N}_{\rm p})}{N-1}\right),
    \label{eq:density}
\end{align}
where $q_0$ denotes the fraction of isolated nodes in the system (see Eq.~\ref{eq:q0_uniform} in Methods~\ref{sec:analytical_derivation_N_M}). If the system is in Regime 1, in which $\kappa$ is constant, the density monotonically approaches $\kappa/4$ as $\Np\to \infty$ (\emph{i.e.}, $q_{0}\to 0$ and $N\to \infty$). On the other hand, if the system is in Regime 2, in which $\Np$ is constant, there is no a \emph{priori} upper bound, and the density exhibits a non-monotonic behaviour. In Regime 2, a change in $\kappa$ has two opposing effects on the network density. First, an increase in $\kappa$ directly increases density through a rise in the probability of edges being created. 
 Second, a shift in $\kappa$ would also increase $N$, which reduces the density through the third term in Eq.~\ref{eq:density}. Since $q_0\to 0$ as $N$ becomes sufficiently large, the latter effect is vanishing, and therefore the density begins to rise with $N$ for a sufficiently large $N$.

\section{Discussion}

Densification and sparsification of networks can occur for two reasons, namely a variation in the population $\Np$ and a variation in the overall activity level $\kappa$. A key finding of this work is that the relative importance of each of these two dynamical factors occasionally change, depending on the social context under study.
By fitting the model to the observed scaling relations, we can detect the main factor that is relevant at a given point in time. Shifts in $\Np$ and/or $\kappa$ affect the activity of all individuals equally, so these parameters could be considered as effective ``temperatures'' of the system. While in this work we studied face-to-face networks of individuals, by its versatility the proposed method could also be used for a wide variety of dynamical systems.

There are some remaining issues for future research. First, the baseline model, a dynamic hidden variable model, relies on a ``homogeneous mixing'' hypothesis, which implies that nodes are connected to each other at random, given their activity levels. If we look at the structural properties of networks, such as triadic closure and community structure, such a hypothesis ---especially for social contexts--- would be unrealistic. However, the fact that the proposed method works remarkably well indicates that, as long as we look at network dynamics at a sufficiently coarse scale, keeping local properties aside, this homogeneous mixing assumption is a good approximation. In fact, introducing a non-random structure would easily make it impossible to obtain analytical expressions that would be needed for identification.
Second, we assumed that the distribution of intrinsic node activities is uniform for simplicity. Ideally, one would need to set this distribution based on empirical evidence. However, measuring the empirical intrinsic activity levels of individuals is extremely difficult because one needs activity levels of totally inactive individuals as well. Furthermore, this parameter might very well have its own temporal evolution. If available, such a rich information would allow for a refinement of the method. Third, while the current method works well for temporal networks whose dynamical regime is occasionally switching, for fixed-regime systems in which the whole dynamics could be explained by either a $\Np$-driven \emph{or} a $\kappa$-driven regime, the proposed regime-switching model is unnecessary. In such cases, one would fit the empirical scaling to each of the two models separately, and then find out which model is better fitted~\cite{Kobayashi2020}.

In many cases, examining the source of network dynamics from the level of each individual would be prohibitively difficult because each individual has his/her own circumstance, and privacy issues often prohibit researchers from obtaining enough information to reveal particular individuals' behaviour. In contrast, global quantities, such as the total numbers of nodes and edges, are much more widely accessible, and therefore utilising these quantities will be inevitable when high-resolution data are difficult to collect. 
A contribution of this work is that the proposed model allows us to detect the role of the two fundamental dynamical factors just by using information on the global network dynamics. Any dynamical processes occurring \emph{on} networks, regardless of whether they are micro- or macro-phenomena, would be largely affected by the underlying dynamics \emph{of} networks. This is in particular the case for spreading processes such as epidemics. A better understanding of the dynamics of densification and sparsification could thus benefit public health policies, which are of central importance for modern social systems.

\section{Methods}

\subsection{Analytical expression for $N$ and $M$}\label{sec:analytical_derivation_N_M}

In this section we derive Eqs.~\eqref{eq:N_main} and \eqref{eq:M_main}.
The numbers of active nodes $N$ and edges $M$ can be expressed as functions of parameters $\kappa$ and $\Np$ (we drop time subscript $t$ for brevity):
\begin{align}
\begin{cases}
    N &= (1- q_0(\kappa,\Np)) \Np,\\
    M &= \frac{\overline{k}(\kappa, \Np) \Np}{2},
\end{cases}
\label{eq:NM_main}
\end{align}
where $\overline{k}(\kappa,\Np)$ denotes the average degree over all the existing nodes including isolated ones, and $q_0(\kappa,\Np)$ denotes the fraction of isolated nodes or equivalently the probability that a randomly chosen node being isolated. 

Let $\rho(a)$ be the density of node activities, and define $u(a,a^\prime)$ as the probability that there is an edge between two nodes having activity levels $a$ and $a^\prime$. The average degree $\overline{k}(\kappa,\Np)$ is given by the number of possible partners times the average of $u(a,a^\prime)$ (see, section~\ref{sec:SI_derivation} in SI for a full derivation):
\begin{align}
    \overline{k}(\kappa,\Np) = (\Np-1) \int \int d a d a^\prime \rho(a)  \rho(a^\prime) u(a, a^\prime),
    \label{eq:k_avg}
\end{align}
It should be noted that Eq.~\eqref{eq:k_avg} is equivalent to the average degree in the standard fitness model~\cite{Boguna2003PRE} if $\Np -1$ is replaced with $N$, which is only asymptotically true in our model.

The fraction of isolated nodes in the system is given by (see, section~\ref{sec:SI_derivation} in SI):
\begin{align}
    q_0(\kappa,\Np) 
    = \int d a^\prime \rho(a^\prime) \left[ 1 - \int u(a^\prime, a) \rho(a) d a \right]^{\Np-1}.
    \label{eq:q_0}
\end{align}
Substituting $\rho(a) = 1$ (\emph{i.e.}, uniform distribution on $[0,1]$) and $u(a, a^\prime) = \kappa a a^\prime$ into Eq.~(\ref{eq:k_avg}) leads to:
\begin{align}
    \overline{k}(\kappa,\Np) 
    = \frac{\kappa }{4} (\Np-1).
\end{align}
Similarly, $q_0$ is given by:
\begin{align}
    q_0(\kappa,\Np) &= \int_0^1   \left( 1 - \frac{\kappa a^\prime}{2}  \right)^{\Np-1}d a^\prime.
\end{align}
By defining a variable $x \equiv 1 - \frac{\kappa a^\prime}{2}$, we have:
\begin{align}
    q_0(\kappa,\Np) &=  \frac{2}{\kappa}\int_{1-\frac{\kappa}{2}}^1  x^{\Np-1} dx  \notag \\
    &= \frac{2}{\kappa\Np}\left[1-\left( 1-\frac{\kappa}{2}\right)^{\Np}\right].
    \label{eq:q0_uniform}
\end{align}
Combining these results with Eq.~(\ref{eq:NM_main}), we have:
\begin{align}
N &= \Np \left[ 1-  \frac{2}{\kappa\Np}\left(1-\left( 1-\frac{\kappa}{2}\right)^{\Np}\right) \right],\label{eq:N}\\
M &= \frac{1}{8} \kappa\Np(\Np-1).\label{eq:M}
\end{align}

It should be noted that if $|1-\kappa/2| < 1$ and $\Np$ is sufficiently large, then $q_0(\kappa,\Np) \simeq 0$ and thereby $N \simeq \Np$ and $M \propto N^2$, as is shown in the study of the static fitness model~\cite{Caldarelli2002PRL,Boguna2003PRE,DeMasi2006PRE}.

\subsection{Bayesian estimation}\label{sec:method_regime_switch}

This section describes how we can infer the model parameters and the dynamical regime at a given time interval $t$. Let ${\rm Pr}(S_t=s|\psi_{t-1};\vect\theta)$ be the probability that a network is in state $s$ (\emph{i.e.}, in Regime $s$) conditional on information available at the end of time interval $t-1$, denoted by $\psi_{t-1}$, for a given set of parameters $\vect\theta=\{\Np,\kappa,\sigma_1,\sigma_2,p_{11},p_{22}\}$.
The likelihood function is then given by:
\begin{align}
L(\{\vect{D}_t\}|\vect{\theta}) = \prod_{t=1}^T \sum_{s=1}^2f(\vect{D}_t|S_t=s,\psi_{t-1};\vect\theta){\rm Pr}(S_t=s|\psi_{t-1};\vect\theta),    
\end{align}
where $\{\vect{D}_t\}$ denotes the sequence of observations $\vect{D}_t = (N_t,M_t)$, and $f$ is given by:
\begin{align}
    f(\vect{D}_t|S_t=s,\psi_{t-1};\vect\theta) = \frac{1}{\sqrt{2\pi\sigma_s^2}}\exp{\left(-\frac{(N_t-h^s)^2}{2\sigma_s^2} \right)}, \;\; s= 1,2.
\end{align}
The log-likelihood function leads to:
\begin{align}
    \log{L}(\{\vect{D}_t\}|\vect{\theta}) &= \sum_{t=1}^T \log\sum_{s=1}^2f(\vect{D}_t|S_t=s,\psi_{t-1};\vect\theta){\rm Pr}(S_t=s|\psi_{t-1};\vect\theta),   \notag \\
    &= \sum_{t=1}^T \log\sum_{s=1}^2\sum_{r=1}^2f(\vect{D}_t|S_t=s,\psi_{t-1};\vect\theta){\rm Pr}(S_{t-1}=r|\psi_{t-1};\vect\theta)p_{rs}.  
\end{align} 
Bayesian inference is conducted based on the relationship $p(\vect{\theta}|\{\vect{D}_t\})\propto L(\{\vect{D}_t\}|\vect{\theta})p(\vect{\theta})$, where $p(\vect{\theta}|\{\vect{D}_t\})$ and $p(\vect{\theta})$ are posterior and prior densities, respectively. For each parameter we collect 20,000 samples (four chains, 5,000 samples after 5,000 burn-in for each chain) generated from the posterior using Markov chain Monte Carlo (MCMC). We implement MCMC using Pystan ver.~2.19.0~\cite{Pystan}, which runs the No-U-Turn sampler (NUTS)~\cite{Nuts2014}. The mean parameter values are summarised in Table~\ref{tab:parameters}. 

\begin{table}[tb]
    \centering
        \caption{Estimated parameters. For each parameter, mean and 95\% credible interval obtained by MCMC are shown at the upper and lower rows, respectively. $N_{\rm max}$ denotes $\max_t\{{N_t}\}$.}
    \begin{tabular}{cccccc}
    \hline
                 & WS-16 & IC2S2-17 & Hospital & Workplace & Prior distribution \\
                \hline
      $\Np$      & 121.152 & 196.236 & 28.087& 109.694  & Uniform($N_{\rm max},2N_{\rm max}$)\\
                 & $[121.005, 121.555]$ & $[196.005,196.884]$ & $[28.002,28.306]$ & $[106.449,112.984]$& \\
      $\kappa$   & 0.467 & 0.222& 0.495& 0.087& Uniform$(0,1)$\\
                 &  $[0.430, 0.512]$ & $[0.215,0.228]$ &$[0.450,0.541]$ & $[0.075,0.101]$&          \\
      $p_{11}$   & 0.930 & 0.969& 0.919& 0.917& Beta$(5,1)$\\
                 &  $[0.853, 0.980]$ & $[0.914,0.996]$ & $[0.857,0.965]$& $[0.836,0.973]$&          \\
      $p_{22}$   & 0.969 & 0.988& 0.926& 0.974&  Beta$(5,1)$\\
                & $[0.932, 0.992]$ & $[0.966,0.999]$  & $[0.861,0.971]$ &  $[0.944,0.993]$&          \\
      $\sigma_{1}$   & 8.587  & 5.358& 1.617& 5.815& Cauchy$(0,2)$\\
                     &$[4.186, 12.618]$  & $[4.360,6.598]$  &$[1.395,1.870]$ & $[4.820,7.064]$&           \\
      $\sigma_{2}$   & 5.828  & 15.776& 1.760& 2.913& Cauchy$(0,2)$\\
                     &$[4.330,7.520]$  & $[14.096,17.714]$ &$[1.532,2.007]$ &  $[2.602,3.270]$ &         \\
      \hline
    \end{tabular}
    \label{tab:parameters}
\end{table}

Now we describe how information is updated in each period. The probability of being in state $s$ conditional on information at time $t$ is written as:
\begin{align}
    {\rm Pr}(S_t=s|\psi_{t};\vect{\theta}) &= \frac{f(\vect{D}_t|S_t=s,\psi_{t-1})\cdot{\rm Pr}(S_t=s|\psi_{t-1})}{\sum_{s}f(\vect{D}_t|S_t=s,\psi_{t-1})\cdot{\rm Pr}(S_t=s|\psi_{t-1})}, \notag \\
    &= \frac{\sum_{r}f(\vect{D}_t|S_t=s,\psi_{t-1})\cdot{\rm Pr}(S_{t-1}=r|\psi_{t-1})p_{rs}}{\sum_{s}\sum_{r}f(\vect{D}_t|S_t=s,\psi_{t-1})\cdot{\rm Pr}(S_{t-1}=r|\psi_{t-1})p_{rs}},
    \label{eq:prob_S_update}
\end{align}
where we drop argument $\vect\theta$ in $f$ for brevity. Given the initial guess for ${\rm Pr}(S_0=r|\psi_{0})$, we can recursively update the probability of being in state $s$.

\subsection{Smoothed probability}\label{sec:smoothed}

The probability ${\rm Pr}(S_t=s|\psi_{t};\vect{\theta})$ obtained in Eq.~\eqref{eq:prob_S_update} is based on information available at time $t$ for a given parameter set $\vect{\theta}$. We can also obtain the probability based on all information, represented by information set $\psi_{T}$.
Let $\vect\xi_{t|T}\equiv [{\rm Pr}(S_t=1|\psi_{T};\vect{\theta}),{\rm Pr}(S_t=2|\psi_{T};\vect{\theta})]^{\prime}$ be the vector of probabilities conditional on information at $T$. $\vect\xi_{t|T}$ can be calculated by conducting backward iteration from $T$~\cite{Hamilton1994book}:
\begin{align}
    \vect\xi_{T-1|T} &= \vect\xi_{T-1|T-1}\odot\{\vect{P}^\prime[\vect\xi_{T|T}(\div)\vect\xi_{T|T-1}]\},\notag \\
    \vect\xi_{T-2|T} &= \vect\xi_{T-2|T-2}\odot\{\vect{P}^\prime[\vect\xi_{T-1|T}(\div)\vect\xi_{T-1|T-2}]\},\notag \\
     &\vdots \notag \\
    \vect\xi_{t|T} &= \vect{\xi}_{t|T-1}\odot\{\vect{P}^\prime[\vect\xi_{t+1|T}(\div)\vect\xi_{t+1|T-1}]\},
\end{align}
where $\odot$ and $(\div)$ denote element-by-element multiplication and  element-by-element division, respectively, and $\vect{P}=(p_{ss})$ is the transition matrix. Note that all the terms in the RHS of the first equality are already known from the previous estimation procedure. After calculating $\vect\xi_{T-1|T}$, we use it to calculate the RHS of the second line. We repeat this until we obtain $\vect\xi_{t|T}$.

\subsection{Validation}\label{sec:validation}
We check the accuracy of the inference method based on synthetic network data generated by the regime-switching hidden variable model. For given parameters $\Np$, $\kappa$, $p_{11}$, and $p_{22}$, and time-varying variables $\{\Npt\}$ and $\{\kapt\}$, we generate sequences of $\{N_t\}$ and $\{M_t\}$ in a way prescribed in the model. When the network at $t$ is in Regime 1 (Regime 2), the true $\Npt$ ($\kapt$) is given by $\Npt = 0.95N_{{\rm p}, t-1}$ ($\kapt=0.95\kappa_{t-1}$), and $\Npt=\Np$ ($\kapt=\kappa$) otherwise. We assume that the initial probability of being in Regime 1 is set at 0.5, and $N_{\rm p{,0}}=\Np$ and $\kappa_{0}=\kappa$. For each parameter, we collect 20,000 samples by MCMC (5,000 samples from four chains after 5,000 burn-in iterations).

The estimated parameters under different sets of ground-truth $\vect\theta$ are summarised in Table~\ref{tab:validation_params} in SI. The estimated smoothed probabilities well match the true states of the generated networks (Fig.~\ref{fig:validation_prob_scatter} in SI). We also group the generated networks based on the probability of being in Regime 1; For each time period, if more than 95\% of the sampled values for ${\rm Pr}(S_t=1|\psi_{t};\widehat{\vect{\theta}})$ are higher (lower) than 0.5, then we classify the corresponding snapshot as being in Regime 1 (Regime 2). If it is not classified as Regime 1 or 2, the network is considered to be in a ``gray area''. As shown in the middle and the right columns of Fig.~\ref{fig:validation_prob_scatter}, the classification of generated networks based on estimated parameters is consistent with the ground truth, while there are some networks that are in gray zones especially when the observed pairs of $(N_t,M_t)$ are overlapped between the two regimes. A comparison between the estimated and the true paths of $\Npt$ and $\kapt$ is also shown in Fig.~\ref{fig:validation_Np_kap}.


\section*{Acknowledgements}
T.K. acknowledges financial support from JSPS KAKENHI Grant nos.~15H05729 and 19H01506. This work was partially supported by the ANR project DATAREDUX (ANR-19-CE46-0008) to M.G.

\section*{Author contributions}
T.K. conceived the research. T.K. and M.G. defined the model. T.K. performed analytical calculation and data analysis. T.K. and M.G. discussed the results and wrote the manuscript.

\clearpage

\setcounter{section}{0}
\setcounter{table}{0}
\setcounter{equation}{0}
\setcounter{figure}{0}
\setcounter{page}{1}
     
\renewcommand{\thetable}{S\arabic{table}}
\renewcommand{\thefigure}{S\arabic{figure}}
\renewcommand{\thesection}{S\arabic{section}}
\renewcommand{\theequation}{S\arabic{equation}}


{\fontsize{18pt}{18pt}\selectfont
 Supporting Information:  \\
 
``The switching mechanisms of social network densification'' \\

\Large{Teruyoshi Kobayashi and Mathieu G\'enois}
\\
}
\vspace{1cm}

\section{Full derivation of Eq.~\eqref{eq:N_main} and \eqref{eq:M_main}}\label{sec:SI_derivation}

We describe a derivation of Eqs.~\eqref{eq:N_main} and \eqref{eq:M_main}.
The numbers of active nodes $N$ and edges $M$ can be expressed as functions of parameters $\kappa$ and $\Np$ (we drop time subscript $t$ for brevity):
\begin{align}
\begin{cases}
    N &= (1- q_0(\kappa,\Np)) \Np,\\
    M &= \frac{\overline{k}(\kappa, \Np) \Np}{2},
\end{cases}
\label{eq:NM_SI}
\end{align}
where $q_{0}$ and $\overline{k}(\kappa,\Np)$ respectively denote the fraction of isolated nodes and the average degree over all the existing nodes including isolated one. 
Let $u(a,a^\prime)$ be the probability that there is an edge between two nodes having activity levels $a$ and $a^\prime$, respectively.
Given the vector of each node's activity $\vect{a} = (a_1, a_2, \ldots, a_{\Np})$, the probability that node $i$ has degree $k_i$ is written as:
\begin{align}
    g(k_i | \vect{a}) &= \sum_{\vect{c}_i} \left[ \prod_{j \neq i} u(a_i, a_j)^{c_{ij}} (1- u(a_i, a_j))^{1-c_{ij}} \right] \delta\left(\sum_{j \neq i} c_{ij}, k_i \right),
\label{eq:g_ki}
\end{align}
where $c_{ij} \in \left\{ 0, 1\right\}$ is the $(i,j)$-element of the $\Np\times\Np$ adjacency matrix, whose $i$th column is given by $\vect{c}_i = (c_{1i}, c_{2i}, \ldots, c_{\Np i})^\top$, and function $\delta(x, y)$ denotes the Kronecker delta.

Let us redefine a product term in the square bracket of \eqref{eq:g_ki} as:
\begin{align}
    f_j(c_{ij}; a_i, a_j) &\equiv u(a_i, a_j)^{c_{ij}} (1- u(a_i, a_j))^{1-c_{ij}}.
\label{eq:def_f}
\end{align}
Since $g(k_i | \vect{a})$ is the convolution of $\left\{ f_j(c_{ij}; a_i, a_j) \right\}_j$, its generating function:
\begin{align}
    \hat{g}_i(z | \vect{a}) \equiv \sum_{k_i} z^{k_i} g(k_i | \vect{a})
\end{align}
is decomposed as:
\begin{align}
    \hat{g}_i(z | \vect{a}) = \prod_{j \neq i} \hat{f}_j(z; a_i, a_j),
\end{align}
where $\hat{f}_j$ is the generating function of $f_j(c_{ij}; a_i, a_j)$, given by:
\begin{align}
    \hat{f}_j(z; a_i, a_j) \equiv \sum_{a_{ij}} z^{a_{ij}} f_j(a_{ij}; a_i, a_j).
    \label{eq:def_fhat}
\end{align}

For a given density of activity levels $\rho(a)$, the degree distribution $p(k_i; \kappa,\Np)$ is defined by the probability that node~$i$ has degree $k_i$:
\begin{align}
    p(k_i ; \kappa,\Np) = \int g(k_i | \vect{a}) \rho(a_1)\rho(a_2)\cdots \rho(a_{\Np})da_{1}da_{2}\cdots da_{\Np},
    \label{eq:p_ki}
\end{align}
A differentiation of $\hat{g}_i(z | \vect{a})$ with respect to $z$ gives us the average degree $\overline{k}(\kappa,\Np)$:
\begin{align}
    \overline{k}(\kappa,\Np) &= \sum_{k_i} k_i p(k_i; \Np) \nonumber\\
    &= \sum_{k_i} k_i \int g(k_i | \vect{a}) \rho(a_1)\cdots \rho(a_{\Np})da_{1}\cdots da_{\Np}, \nonumber\\
    &= \frac{d}{dz} \int \hat{g}_i(z | \vect{a})\rho(a_1)\cdots \rho(a_{\Np})da_{1}\cdots da_{\Np} \Bigr|_{z=1} \nonumber\\
    &= \frac{d}{dz} \int \rho(a_i) da_i \prod_{j \neq i} \int \hat{f}_j(z; a_i, a_j) \rho(a_j) da_j \Bigr|_{z=1} \nonumber\\
    &= \int \rho(a_i) da_i \frac{d}{dz} \left[ \int \hat{f}(z; a_i, h) \rho(h) dh \right]^{\Np-1} \Bigr|_{z=1} \nonumber\\
    &= (\Np-1)  \int \rho(a_i) d a_i \left[ \int d a \rho(a) \hat{f}(z; a_i, a) \right]^{\Np-2} \int d a \rho(a) \frac{d}{dz} \hat{f}(z; a_i, a)\Bigr|_{z=1}.
    \label{eq:k1}
\end{align}

From Eqs.~(\ref{eq:def_f}) and (\ref{eq:def_fhat}), we have $\hat{f}(z; a_i, a) = \sum_{c_{ij}} z^{c_{ij}} f(c_{ij}; a_i, a) = (z-1)u(a_i, a) + 1$. It follows that:
\begin{align}
    &\int da \rho(a) \hat{f}(z; a_i, a) = (z-1) \int da \rho(a) u(a_i, a) + 1,\\
    &\int da \rho(a) \frac{d}{dz} \hat{f}(z; a_i, a) = \int da \rho(a) u(a_i, a).
\end{align}

Substituting these equations into Eq.~\eqref{eq:k1} leads to:
\begin{align}
    \overline{k}(\kappa,\Np) = (\Np-1)\int \int d a d a^\prime \rho(a)  \rho(a^\prime) u(a, a^\prime).
\end{align}

From \eqref{eq:p_ki}, the probability of a node being isolated, $q_0(\kappa,\Np) \equiv p(k_i = 0; \kappa,\Np)$, is given by:
\begin{align}
    q_0(\kappa,\Np) 
    &= \int g(k_i =0 | \vect{a})\rho(a_1)\cdots \rho(a_{\Np})da_{1}\cdots da_{\Np}, \nonumber\\
    &= \int d a_i \rho(a_i) \left[ 1 - \int u(a_i, a) \rho(a) d a \right]^{\Np-1}.
    \label{eq:q_0}
\end{align}
Then, substituting $\rho(a) = 1$ (i.e., uniform distribution on $[0,1]$) and $u(a, a^\prime) = \kappa a a^\prime$ into Eq.~(\ref{eq:k_avg}) gives:
\begin{align}
    \overline{k}(\kappa,\Np) 
    = \frac{\kappa }{4} (\Np-1).
\end{align}
Similarly, substituting the same conditions into Eq.~(\ref{eq:q_0}) gives:
\begin{align}
    q_0(\kappa,\Np) &= \int_0^1   \left( 1 - \frac{\kappa a_i}{2}  \right)^{\Np-1}d a_i.
\end{align}

By defining a variable as $x \equiv 1 - \frac{\kappa a_i}{2}$, we have:
\begin{align}
    q_0(\kappa,\Np) &=  \frac{2}{\kappa}\int_{1-\frac{\kappa}{2}}^1  x^{\Np-1} dx  \notag \\
    &= \frac{2}{\kappa\Np}\left[1-\left( 1-\frac{\kappa}{2}\right)^{\Np}\right].
\end{align}
Note that $q_0(\kappa,1)=1$ and $\lim_{\Np\to \infty}q_0(\kappa,\Np)=0$.
Combining these results with Eq.~(\ref{eq:NM_SI}), we have:
\begin{align}
\begin{cases}
    N &= \Np \left[ 1-  \frac{2}{\kappa\Np}\left(1-\left( 1-\frac{\kappa}{2}\right)^{\Np}\right) \right],\\
    M &= \frac{1}{8} \kappa\Np(\Np-1).
\end{cases} \label{eq:SI_NM}
\end{align}

\clearpage

\begin{table}[tbh]
    \centering
        \caption{Estimated parameters based on synthetic networks. For each parameter, mean and 95\% credible interval obtained by MCMC are shown at the upper and lower rows, respectively. Parameters without hats, annotated in the first row and the first column, denote true values used in the generation of synthetic networks.}
    \begin{tabular}{ccccc}
    \hline
    
        $(\Np,\kappa)$         & $(100,0.2)$ & $(100,0.4)$ & $(200,0.2)$ & $(200,0.4)$  \\
                \hline
      $\Nph$      & 100.900 & 100.208  & 199.447 & 199.780  \\
                 & $[99.912,101.919]$ & $[99.493,100.909]$ & $[198.521, 200.381]$ & $[199.135,200.417]$\\
      $\widehat{\kappa}$ & 0.203 &0.391 & 0.197& 0.406\\
                 &  $[0.195,0.211]$ & $[0.374,0.408]$ &$[0.192,0.201]$ & $[0.397,0.416]$           \\
      $\widehat{p}_{11}$   & 0.924 & 0.950& 0.927& 0.913\\
        $(p_{11}=0.95)$         &  $[0.856,0.972]$ & $[0.882,0.989]$ & $[0.869,0.970]$& $[0.836,0.966]$           \\
      $\widehat{p}_{22}$   & 0.951 & 0.980& 0.938& 0.958 \\
        $(p_{22}=0.95)$        & $[0.907,0.982]$ & $[0.952,0.996]$  & $[0.885,0.976]$ &  $[0.919,0.984]$          \\
      \hline
    \end{tabular}
    \label{tab:validation_params}
\end{table}

\begin{figure*}[tbh]
    \centering
    \includegraphics[width=16.5cm]{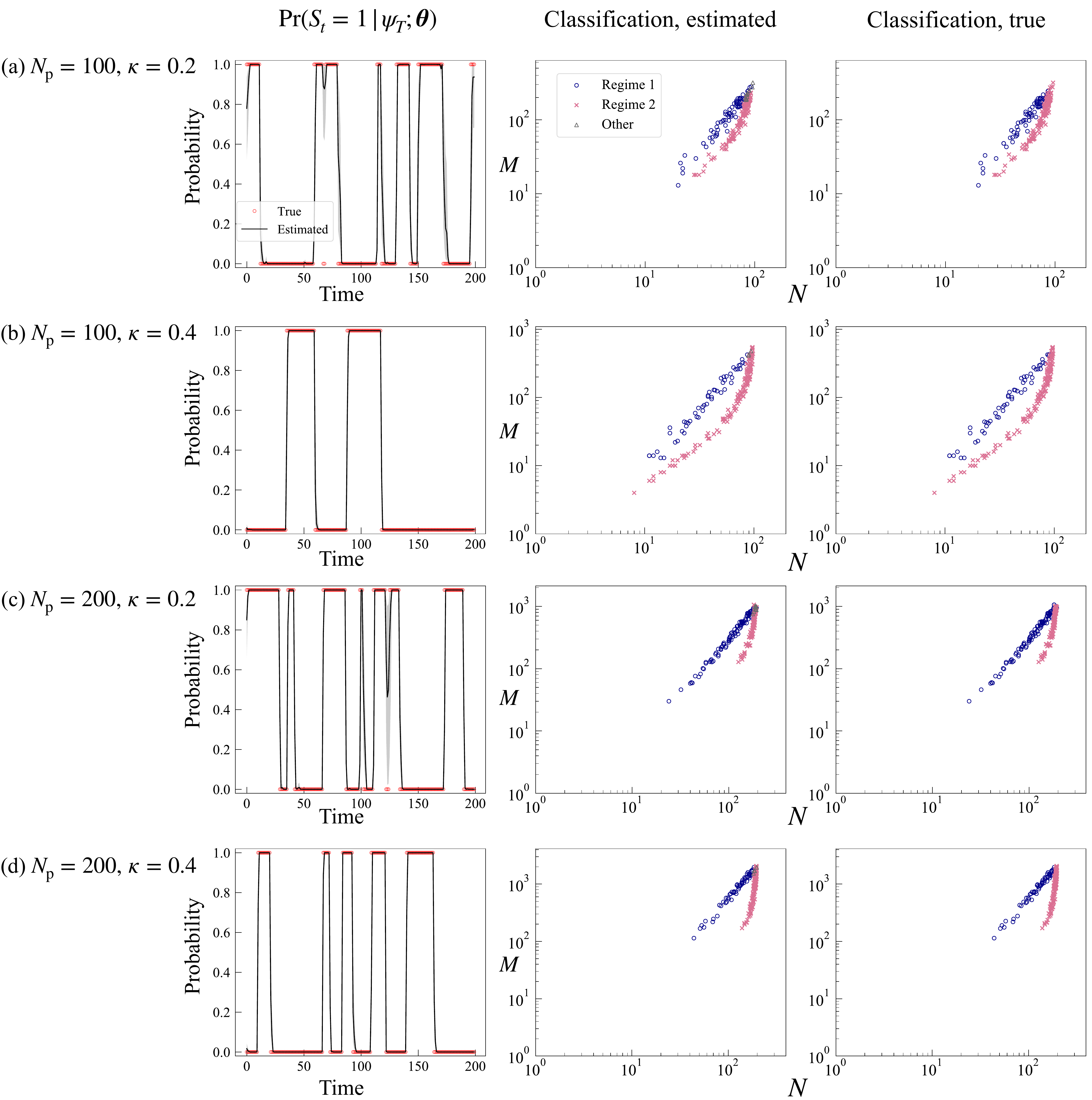}
    \caption{Validation of the classification of regimes. In the left column, the mean of 20,000 sampled values for $\text{Pr}(S_t=1|\psi_T;\vect{\widehat\theta})$ and their 95\% credible intervals are denoted by black solid and gray shading, respectively, and the ground-truth is denoted by red circle. In the middle and right columns, estimated and true classification of network regimes are respectively shown. Each dot corresponds to a generated network. If more than 95\% of sampled probabilities, ${\rm Pr}(S_t=1|\psi_{t};\widehat{\vect{\theta}})$, are higher (lower) than 0.5, then we classify the network as being in Regime 1 (Regime 2). If it is not classified as Regime 1 or 2, the network is considered to be in a ``gray area''.}
    \label{fig:validation_prob_scatter}
\end{figure*}

\begin{figure*}[tbh]
    \centering
    \includegraphics[width=16.5cm]{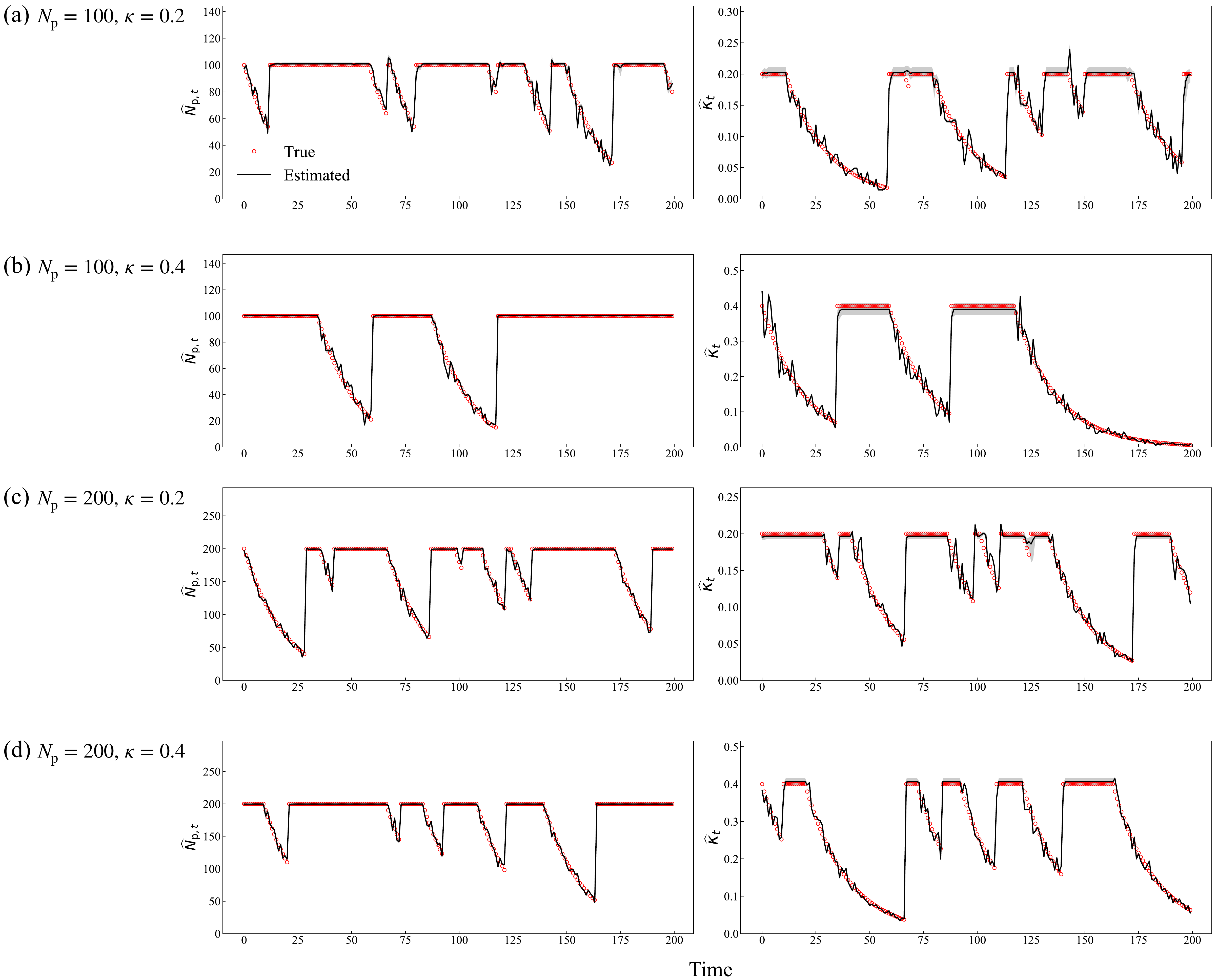}
    \caption{Validation of estimated parameters. True parameter value is denoted by red circle, and the mean of 20,000 parameter sampling and their 95\% credible intervals are denoted by black solid and gray shading, respectively.}
    \label{fig:validation_Np_kap}
\end{figure*}

\begin{figure*}[tb]
    \centering
    \includegraphics[width=15.5cm]{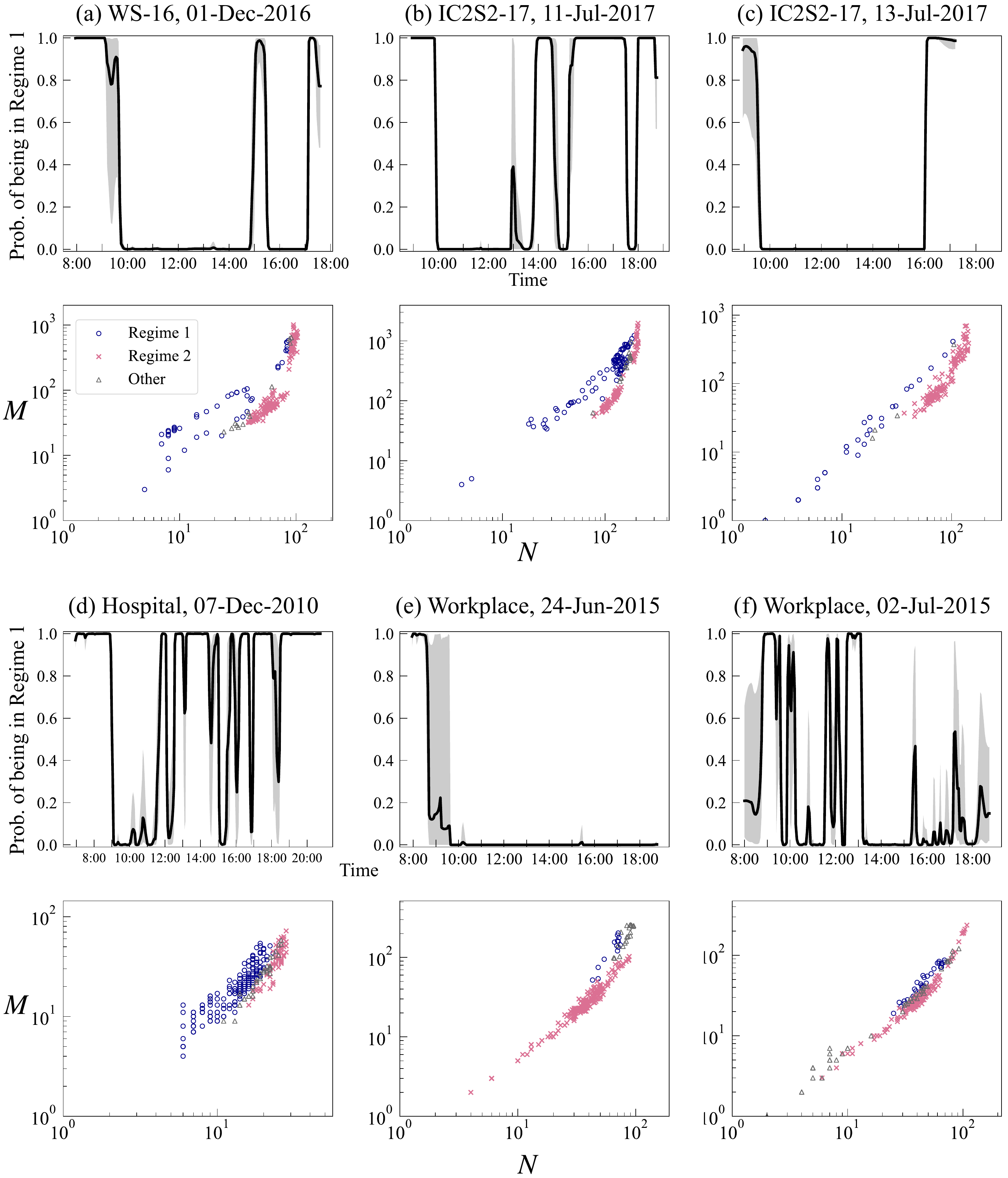}
    \caption{Identification of the dynamical regime. Upper panels show the smoothed probability of being in Regime 1 (\emph{i.e.}, $\Np$-driven dynamics) at each time window. 95\,\% credible interval is indicated by shading. Lower panels show $N$-$M$ plots with classified regimes being denoted by different colours and symbols.}
    \label{fig:SI_regime_NM_result}
\end{figure*}

\begin{figure*}[tb]
    \centering
    \includegraphics[width=16.5cm]{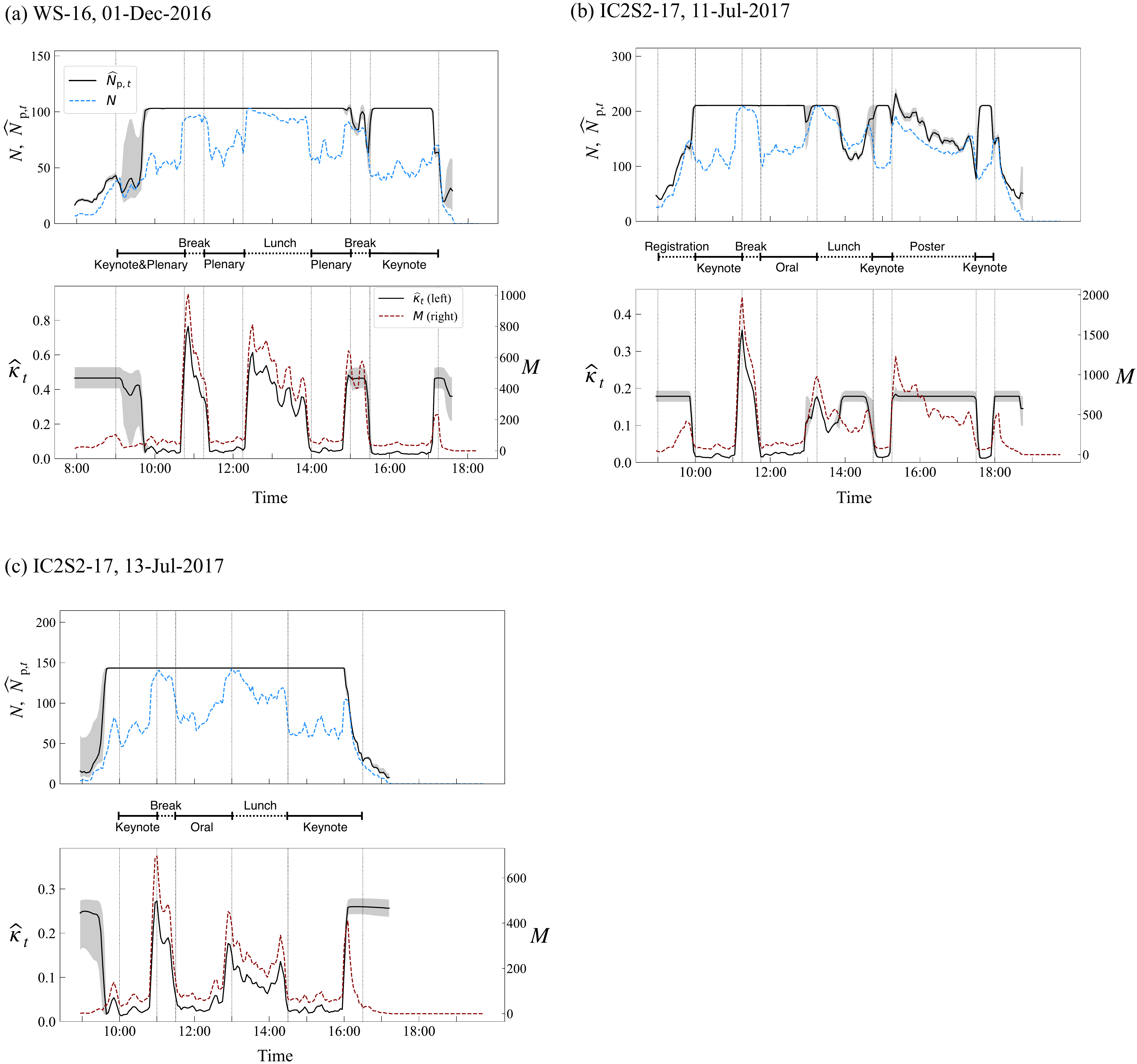}
    \caption{Estimation of $\Npt$ and $\kappa_t$ for (a) WS-16 (day 2), (b) IC2S2-17 (day 2) (c) IC2S2-17 (day 4). $\Npht$ and $\kapht$ are shown in the upper and the lower panels, respectively, and 95\% credible interval is indicated by shading. In the middle, the official conference schedule of the day is shown.}
    \label{fig:SI_conf_schedule}
\end{figure*}

\begin{figure*}[tb]
    \centering
    \includegraphics[width=16.5cm]{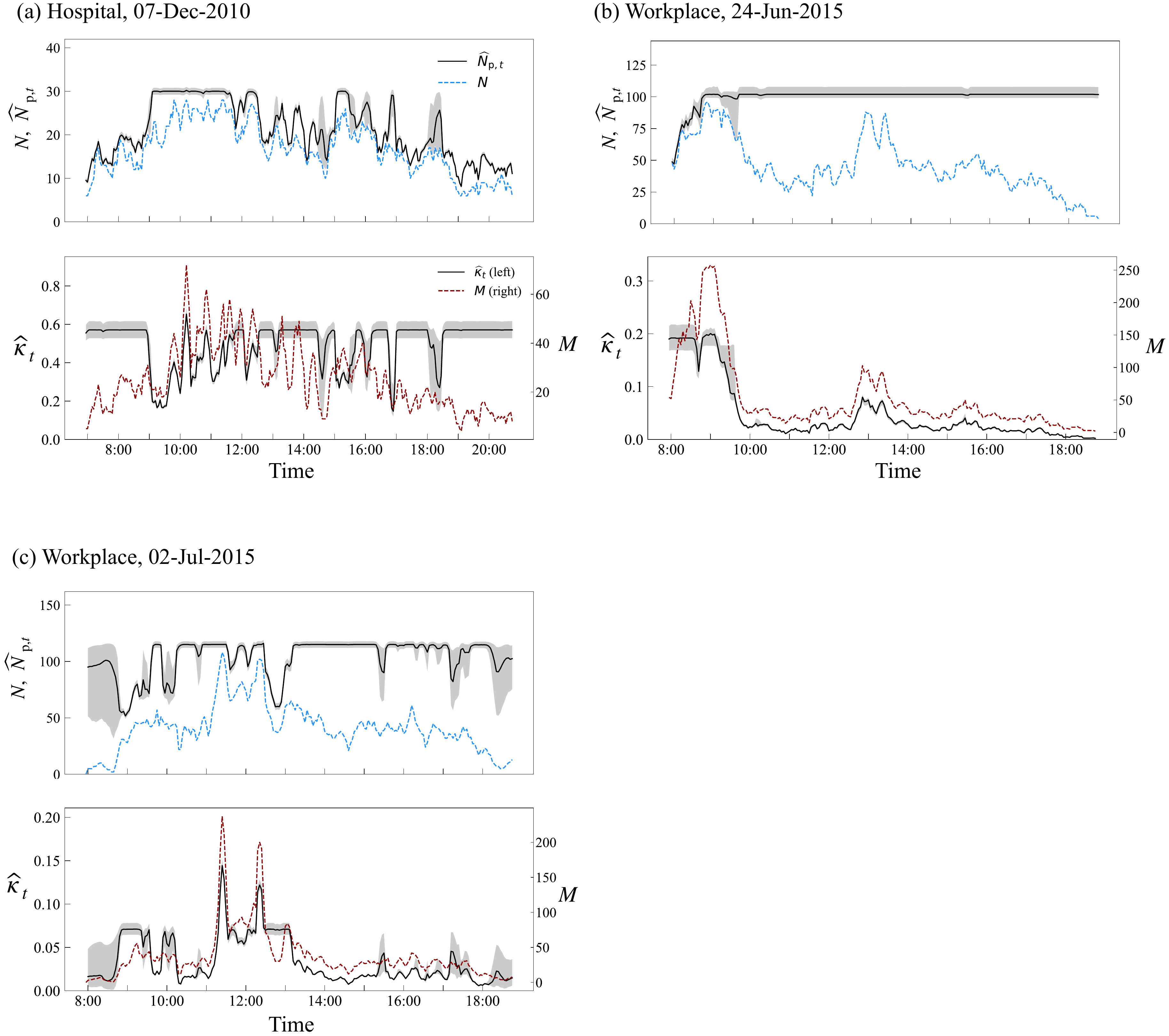}
    \caption{Estimation of $\Npt$ and $\kappa_t$ for (a) Hospital and (b) and (c) Workplace data. $\Npht$ and $\kapht$ are shown in the upper and the lower panels, respectively, and 95\% credible interval is indicated by shading.}
    \label{fig:SI_HosWork_schedule}
\end{figure*}

\end{document}